\documentclass[usenatbib,useAMS,usegraphicx]{mn2e}
\usepackage{longtable}

\title[Australia Telescope Low-brightness Survey]
{ATLBS: the Australia Telescope Low-brightness Survey}

\author[Subrahmanyan, Ekers, Saripalli \& Sadler]
{R. Subrahmanyan,$^{1,2}$\thanks{email: rsubrahm@rri.res.in}
 R. D. Ekers,$^{2}$ 
 L. Saripalli,$^{1,2}$ 
 and E. M. Sadler$^{3}$ \\
$^1$Raman Research Institute, C. V. Raman Avenue, Sadashivanagar, Bangalore
560080, India \\ 
$^2$Australia Telescope National Facility, PO Box 76, Epping NSW 1710,
Australia \\ 
$^3$School of Physics, University of Sydney NSW 2006, Australia}

\begin{document}

\maketitle

\begin{abstract}
We present a radio survey carried out with the Australia Telescope Compact
Array.  A motivation for the survey was to make a complete inventory of the
diffuse emission components as a step towards a study of the cosmic evolution in radio
source structure and the contribution from radio-mode feedback on
galaxy evolution.  The Australia Telescope low-brightness survey (ATLBS)
at 1388~MHz covers 8.42~deg$^{2}$ of the sky in an observing mode designed to
yield images with exceptional surface brightness sensitivity and low confusion. The
survey was carried out in two adjacent regions on the sky centred at
RA: $00^{\rm h}~35^{\rm  m}~00^{\rm s}$, DEC: $-67\degr~00\arcmin~00\arcsec$ 
and RA: $00^{\rm h}~59^{\rm m}~17^{\rm s}$, DEC: $-67\degr~00\arcmin~00\arcsec$ (J2000.0).
The ATLBS radio images, made with 0.08~mJy~beam$^{-1}$ rms noise and
$50\arcsec$ beam, detect a total of 1094 sources with peak flux exceeding 0.4~mJy~beam$^{-1}$.
The ATLBS source counts were corrected for blending, noise bias, resolution, and
primary beam attenuation; the normalized differential source counts are
consistent with no upturn down to 0.6~mJy. The percentage integrated
polarization $\Pi_{0}$ was computed after corrections for the polarization bias in
integrated polarized intensity; $\Pi_{0}$ shows an increasing trend with
decreasing flux density. Simultaneous visibility measurements made with longer baselines
yielded images, with $5\arcsec$ beam, of compact components in sources
detected in the survey. The observations provide a measurement of the
complexity and diffuse emission associated with mJy and sub-mJy radio
sources. 10\% of the ATLBS sources have more than half of their flux density in
extended emission and the fractional flux in diffuse components does not
appear to vary with flux density, although the percentage of sources that
have complex structure increases with flux density. The observations are consistent with
a transition in the nature of extended radio sources 
from FR-{\sc ii} radio source morphology, which dominates the mJy
population, to FR-{\sc i} structure at sub-mJy flux density.  
\end{abstract}

\begin{keywords}
techniques: interferometric -- surveys -- galaxies: active -- galaxies: nuclei
--  galaxies: evolution -- galaxies: high-redshift -- radio continuum:
galaxies
\end{keywords}

\section{Introduction}

Our understanding of galaxy evolution across cosmic time depends on
multi-wavelength data, which are products of ultra-deep surveys across the
electromagnetic spectrum that have been made using the most sensitive
observatories in operation. Instrument design constraints and resource
limitations usually lead to survey strategies that range from all-sky surveys
with low angular resolution and sensitivity to ultra-deep surveys of small sky
regions that are made with high angular resolution; different survey 
strategies address different components of source populations and different
aspects of galaxy evolution.  

The radio component of these multi-wavelength
campaigns that target small sky regions has most often been done 
using interferometer telescopes configured to give images with 
sub-arcsec resolution that are comparable to or better than the corresponding
optical surveys. These radio surveys usually have extremely good flux sensitivity and
are capable of detecting $\mu$Jy emission from distant galaxies. However, Fourier
synthesis imaging that is done with widely spaced
interferometer elements---in order to image with high angular
resolution---tend to lack surface brightness sensitivity, which
is the ability to detect faint extended emission components. This is partly
due to missing short spacings, which implies missing information on extended
emission, and partly because of incomplete visibility
coverage, which results in increased confusion. Furthermore, 
imaging with an interferometer array that has complete visibility coverage will
still have less (redundant) short spacings than a filled aperture and hence less
sensitivity to extended structure. Consequently, the
ultra-deep radio images might fail to reproduce extended emission components
associated with galaxies, although extremely faint compact components are 
represented in the images. 

The normalized differential radio source counts are observed to show an upturn
below flux density of about 1~mJy \citep{wi85} indicating a rapid increase in
the number of faint sources at these flux density levels, which might
constitute a new population.  The bulk of these faint radio sources in the
0.1--1~mJy range in flux density are identified with early type galaxies
\citep{ma08}, with radio structure believed to be of the FR-{\sc i}
\citep{fr74} type \citep{pa07}, which often have associated extended emission
components. Therefore a complete census of the radio emission associated with
faint 
galaxies, at these flux density levels and at intermediate redshifts of
$z=1$--2, requires  
imaging with good surface brightness sensitivity. Additionally, 
the radio morphology of the relatively lower
surface brightness extended emission is a clue to the nature of the radio
source and is a means of distinguishing between active galactic nuclei (AGNs)
of different classes.   

The Square Kilometre Array (SKA\footnote{http://www.skatelescope.org}) has the
potential to detect AGNs, including the radio quiet population, out to
redshift $z \approx 6$. As compared to X-ray and optical surveys, radio
imaging with the next generation telescopes are likely to become the instrument of choice
for identifying high redshift AGNs, in particular the obscured AGNs
\citep{ja04}. In this context, it is important to quantify the expectations for
radio source confusion from both compact and extended radio source components
at faint flux density levels as this is a potential 
limitation to imaging with sub-$\mu$Jy sensitivity. A characterization of the radio
sky at faint flux density---in total intensity and polarization---is a useful
input to simulations of the radio sky and optimization of SKA
array configurations as well as observing strategies.

The survey presented here has been made with the specific goal of providing a
view of the low surface brightness radio emission associated with mJy and
sub-mJy radio sources, to provide a database for their characterization,
assessment of the cosmic evolution of extended radio components and their
influence on galaxy evolution. The surface brightness sensitivity of this
survey, which we refer to as the Australia Telescope Low-Brightness Survey
(ATLBS), is about a factor of five better than 
any previous survey with comparable resolution \citep{sub07}.
In this first paper we present the ATLBS survey together 
with the source counts and a population study of the radio structures.
Forthcoming papers will present 
detailed radio structures, optical identifications, polarization analysis, and
explore in depth the open problems like, for example, cosmic evolution of low
power radio galaxies, evolution of the radio source structure with
cosmic epoch and the role of kinetic-mode feedback from AGNs on galaxy
environment and galaxy evolution, where progress depends on our understanding
of the low surface brightness radio sky.

\section{Survey strategy}

Deep surveys that target weak extended emission components and attempt to get
close to the confusion limit require good control of systematics.
Interferometers are preferred over single-dish
scanning surveys because of the inherent insensitivity to the mean sky
background and, consequently,
the vastly superior stability in the zero-point level that is achievable 
in images constructed using Fourier synthesis techniques.

Extended radio sources usually have steep spectral indices and, therefore, it
has often been argued that surveys for the detection of low surface
brightness sources ought to be made at relatively low radio frequencies. While
this argument is correct, there are practical issues that merit consideration. 
At frequencies below about 1~GHz, elements forming 
interferometer telesopes often have poorly
defined fields of view (primary beams), relatively larger sidelobe levels, and
confusion arising from radio sources limits the attainable dynamic range.
Additionally, the ionosphere introduces time-varying phase errors that are
difficult to calibrate and correct.  The available bandwidth is also limited
at low frequencies. For these reasons, the optimum frequency for deep surveys
with high surface brightness over moderate sky areas is perhaps around 1~GHz
today, and might move to lower frequencies as technology and calibration
algorithms relevant for wide-field imaging at low radio frequencies improve.

High fidelity surveys for extended sources with low surface brightness
requires good spatial frequency coverage.  Holes in the visibility plane---the
{\it uv}-coverage---effectively reduces the number of independent synthesized beam
areas within the primary beam and, consequently, sidelobe confusion due to discrete
sources limits the image dynamic range and quality.  Most 2D Fourier synthesis
imaging arrays like, for example, the Very Large Array (VLA) and the Giant
Metrewave Radio Telescope (GMRT) have array configurations optimized for
imaging performance in snap-shot mode and in cases where most of the sky
region imaged is empty.  The deconvolution algorithms implemented in software
packages used in processing the data from such arrays also implicitly assume
that most of the sky is empty. However, deep surveys that attempt to get close
to sidelobe-confusion limits require filled {\it uv}-coverage and this motivation has led us
to use the EW Australia Telescope Compact Array (ATCA) for the observations
presented here.

The ATCA has five movable antennas on a 3-km EW railtrack. We
have used the array to image fields using four 750-m array
configurations---the 750A, 750B, 750C and 750D arrays.  Together, they
provide $4 \times 10 = 40$ baselines and because the ATCA antennas are 22~m in
diameter, the 40 spacings provide a nearly complete coverage
over the 0--750~m range.  At the 750-m baseline, Earth rotation would move the
visibility measurement point through a spatial wavelength corresponding to the
antenna diameter in about 7~min.  Therefore, the observing strategy was to
mosaic image 19 distinct antenna pointing positions during a single observing
session of 12~hr, cycling through all the pointing positions and observing each
for 20~sec so that all the 19 positions would be re-visited once every
7~min.  This ensured that the elliptic {\it uv}-tracks have complete azimuthal
coverage for each pointing. Observations with the ATCA were made using the
20-cm band.

The 19 pointing positions observed during any
observing session were arranged to tile the sky in a hexagonal pattern 
so as to cover the sky with
a smaller number of pointings compared to a square grid. 
In the 20-cm band, the ATCA antenna primary beam has a full width at half
maximum (FWHM) of about $35\arcmin$, and mosaic imaging of large angular scale
extended structure requires a sky plane `Nyquist' sampling interval of
$19\farcm5$. However, we have adopted to survey the sky as a collage of
individual image tiles without attempting to image structure on the scale of
the primary beam or greater; therefore, the hexagonal-packed adjacent
pointings are spaced 
$28\farcm6$ and this spacing is sufficient to cover the sky with fairly
uniform sensitivity.  The sequence in which the pointings were observed 
was selected to minimize time lost while the telescope cycled through the
pointings. 

\section{The mosaic observations}

\begin{table}
\caption{Journal of the ATCA radio observations.}
\begin{tabular}{@{}rcr}
\hline
Survey region & Array configuration & Date \\
\hline
A & 750A & February 26 2004 \\
 & 750B & January 12 2005 \\
 & 750C & November 12 2004 \\
 & 750D & July 02 2004 \\
B & 750A & February 28 2004 \\
 & 750B & January 13 2005 \\
 & 750C & November 11 2004 \\
 & 750D & July 03 2004 \\
\hline
\end{tabular}
\end{table}

Since our observations use the ATCA with antennas on EW baselines, the
survey region was constrained to be at high southern declinations and far from
the 
equator, in order to image with close to circular synthesized beams. High
southern declinations are also preferred so that the fields might be far from
the Sun, and Solar interference would be minimized at epochs when the
fields are scheduled for day-time observing.  To avoid
shadowing at the shortest 30-m baseline, the field centres had to be south 
of $-50\degr$ declination. On the other hand, since follow-up optical
observing with existing southern telescopes are important for the science
goals,  
low declinations were preferred so that optical observing could be 
through low airmass.  Since the ATCA is located at a latitude of $-30\degr$,
regions at very high southern declinations were avoided so that the survey
might be made at reasonably high telescope elevations, avoiding problems that
might arise from ground spillover in the antenna radiation pattern.
High Galactic latitudes were preferred since the
background sky brightness and hence the system temperature would be lower. 

A 10\% departure from circularity in the synthesized beam was considered acceptable, and a
pair of sky regions were selected at $-67\degr$ declination after examining
these declination strips in the Sydney University Molonglo Sky Survey (SUMSS;
\citet{bo99}) and in the Parkes-MIT-NRAO survey (PMN; \citet{gr93}) 
and ensuring that they were relatively devoid of strong sources.
The sky regions selected for the mosaic imaging survey, which we refer to as
ATLBS survey region A and B, have field centres at coordinates (J2000.0 epoch)
RA: $00^{\rm h}~35^{\rm  m}~00^{\rm s}$, DEC: $-67\degr~00\arcmin~00\arcsec$ 
and RA: $00^{\rm h}~59^{\rm m}~17^{\rm s}$, DEC:
$-67\degr~00\arcmin~00\arcsec$ respectively.  The two 
regions are individually mosaics that are covered using 19 pointings, and they
are located beside each other on the sky.  

The observations of each of these two sky regions were made in the four 750-m
arrays and each of these four sessions were of 12-hr duration (time shared
between the 19 pointing positions). The 20-cm band data were acquired in two
128-MHz wide bands centred at 1344 and 1432~MHz. Each band was
covered by 16 independent frequency channels.  Multi-channel continuum
visibility data were accumulated in full polarization mode: the ATCA antennas
have feeds with linear polarization outputs labeled X and Y and the
polarization products XX, YY, XY and YX are accumulated by the correlator.  
A journal of the radio observations is in Table~1.

The ATCA has six antennas: the location of the sixth
antenna---ca06---is 
fixed and provides baselines between 3 and 5~km with the other five
antennas in the 750~m arrays.  The resulting {\it uv}-coverage in our observations
is completely filled out to 750~m and is sparsely covered in the 3--5~km
range; there is a significant `hole' in the coverage between 750~m and 3~km.

\section{Radio imaging}

\begin{figure*}
\includegraphics[width=177mm]{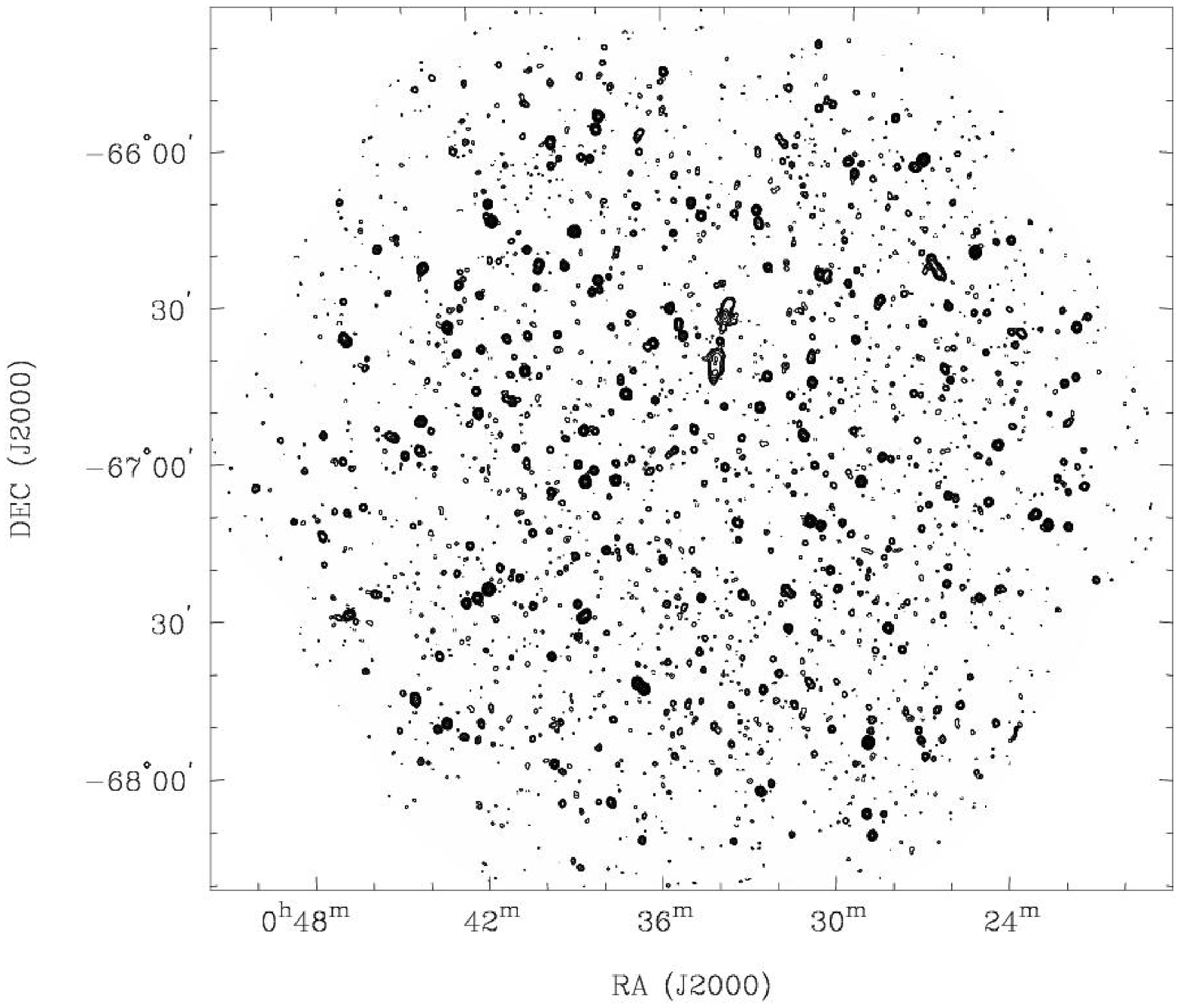}
\caption{Mosaic image of ATLBS survey region A made with a beam of FWHM $52\farcs4
  \times 47\farcs4$ at P.A. $6\degr$. Contour levels are at $-$0.25, 0.25,
  0.5, 1.0, 2.0, 4.0, 8.0, 16.0, 32.0, 64.0, 128.0 and 256.0~mJy~beam$^{-1}$.}
\end{figure*}

\begin{figure*}
\includegraphics[width=177mm]{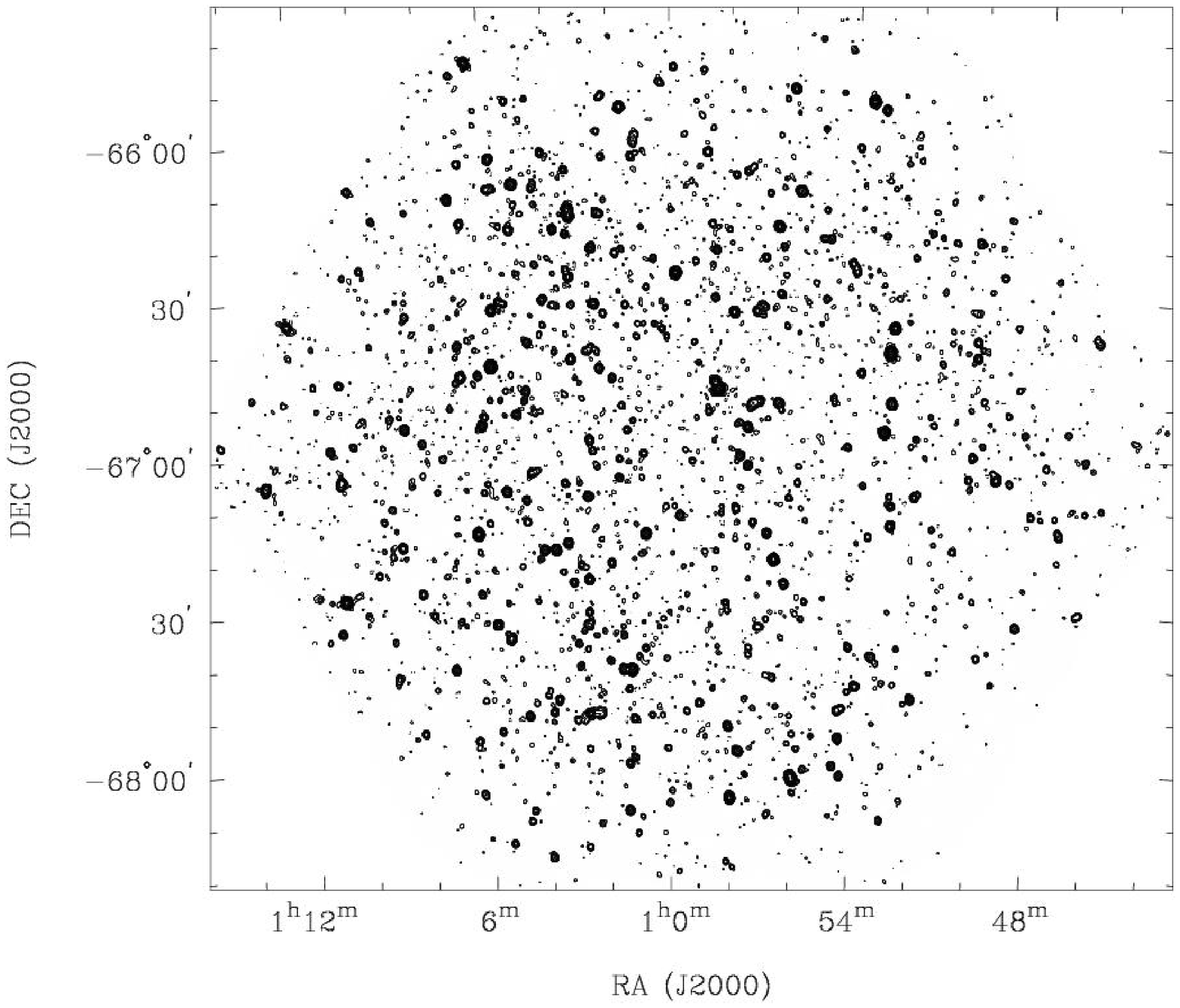}
\caption{Mosaic image of ATLBS survey region B made with a beam of FWHM $52\farcs8
  \times 47\farcs4$ at P.A. $7\fdg3$. Contour levels are at $-$0.25, 0.25,
  0.5, 1.0, 2.0, 4.0, 8.0, 16.0, 32.0, 64.0, 128.0 and 256.0~mJy~beam$^{-1}$.}
\end{figure*}

The data were processed and imaged using the radio interferometer data
reduction package MIRIAD. The calibrator visibilities were 
first viewed using visualization
tools in MIRIAD and obviously erroneous data were rejected.
The multi-channel continuum visibility data were calibrated in amplitude,
phase and for the bandpass using periodic observations of the unresolved
calibrator PKS~B2353$-$686; 
the absolute flux density scale was set using observations of the primary
calibrator PKS~B1934$-$638.
Polarization calibration for the telescope were derived from the full 
polarization products measured on PKS~B2353$-$686. A first pass was made on
rejecting data with interference using an automated algorithm that examined
the Stokes~V visibilities and rejected the data corresponding to all
polarization products at the same times and channels where the Stokes~V
visibility amplitudes exceeded a threshold of 4 times the rms thermal noise.
The visibility data on the survey fields in individual baselines, in 
XY and YX polarization products, were visualized as grey-scale displays of
channel versus time, and obviously discordant data values were rejected.  The
data in XX and YY polarization products were also rejected during these times
and for the same channels.  

The mosaicing strategy adopted here is to individually image and deconvolve
the different pointings and then combine them to produce a single wide-field
image.  This approach---as compared to a `joint' approach wherein all pointings
are handled simultaneously during the imaging and deconvolution steps---is
appropriate in the present case where dynamic ranges exceeding several hundred
or so are 
desired and imaging structures on the scale of the antenna primary beam
and larger is not a requirement.

\subsection{A low resolution image with high surface brightness sensitivity}

The individual pointing visibilities were separately processed; ca06 was
excluded from this initial analysis that was aimed at making Stokes~I images
with high surface brightness sensitivity using the 0--750~m baselines. As a
first step images of $4\degr \times 4\degr$ were constructed, which were 
seven times wider than the primary beam FWHM, so that sources in the first
sidelobe would be represented. The wide-field images were deconvolved, the
`clean' components representing sources within the main lobe of the primary
beam were isolated and their contribution to the visibility subtracted, and
then images were constructed representing the contributions from sources in the
sidelobes.  These features are significantly different from the point
spread function (synthesized beam) due to azimuthal asymmetries in the
sidelobe pattern together with the alt-azimuth nature of the mounts of the
ATCA antennas. The response to sources in the sidelobes were deconvolved and
represented as `clean' components---composed of positive and negative
components---and this model representing all of the response to sources
outside the main lobe of the primary beam was then subtracted from the 
multi-channel visibilities.  In the subsequent reduction, only the primary
beam main lobe area was considered. The visibility data were then imaged, 
deconvolved and self-calibrated iteratively. Initially the phases alone
were self-calibrated, then the amplitudes in the two frequency bands were
allowed to separately scale (in an amplitude self-calibration step); this
effectively amounts to an amplitude correction based on the weighted mean
spectral index of all of the sources in the field. In final iterations the
visibilities were self calibrated in amplitude and phase.  The processed
visibilties were used to image the individual fields adopting the
multi-frequency deconvolution algorithm \citep{sa94}; this allows for
differing spectral indices among the sources in the field and also corrects,
to first order, for effects arising from the variation in the primary beam
with frequency. 

The images corresponding to the 19 pointings constituting each field were
lastly combined as a linear mosaic, fully correcting for the primary beam
attenuation within the mosaic region but retaining attenuations at the edges
of the mosaic to prevent excessive noise amplification there. These mosaic
images of the ATLBS regions A \& B are shown in Figs. 1 \& 2 respectively. The
images have 
an rms noise of $\sigma = 85~\mu$Jy~beam$^{-1}$ and the lowest contour level
displayed 
is at $3 \sigma$.  The peak in the image of region A is
104~mJy~beam$^{-1}$ and that in region B is 246~mJy~beam$^{-1}$. The negative
peaks in the two images are at levels $-450$ and $-575~\mu$Jy~beam$^{-1}$
respectively, which are at the 5--7$\sigma$ level. The ratio of
image peak to rms noise exceeds 1000 and the images show no
obvious artefacts due to calibration or imaging errors above a level of $3
\sigma$.

Images in Stokes~Q and U were constructed using the same
baselines and weighting schemes used in making the Stokes~I images.
In making these images, we followed the procedure of first constructing
wide-field $4\degr \times 4\degr$ images, isolating the clean components
representing emission from the side lobes of the primary beam, subtracted these
components from the visibility data, then constructed images representing
emission from the main lobes corresponding to each pointing position.  The
images corresponding to the different pointings were then combined as a linear
mosaic to reconstruct the Stokes~Q and U emission from ATLBS region A and,
separately, region B. 

Additionally, we have constructed Stokes~V images of the two fields. No
obvious sources are seen above the image thermal noise.  The peak-to-peak
intensity fluctuations in the Stokes~V images are in the range $\pm
0.47$~mJy~beam$^{-1}$, at $\pm 5.5 \sigma$; these values are similar in
magnitude to the negative peaks in the Stokes~I images indicating that the
negative peaks in the Stokes~I images are consistent with the image thermal noise.

\subsection{A high resolution image of the compact components}

\begin{figure*}
\includegraphics[width=177mm]{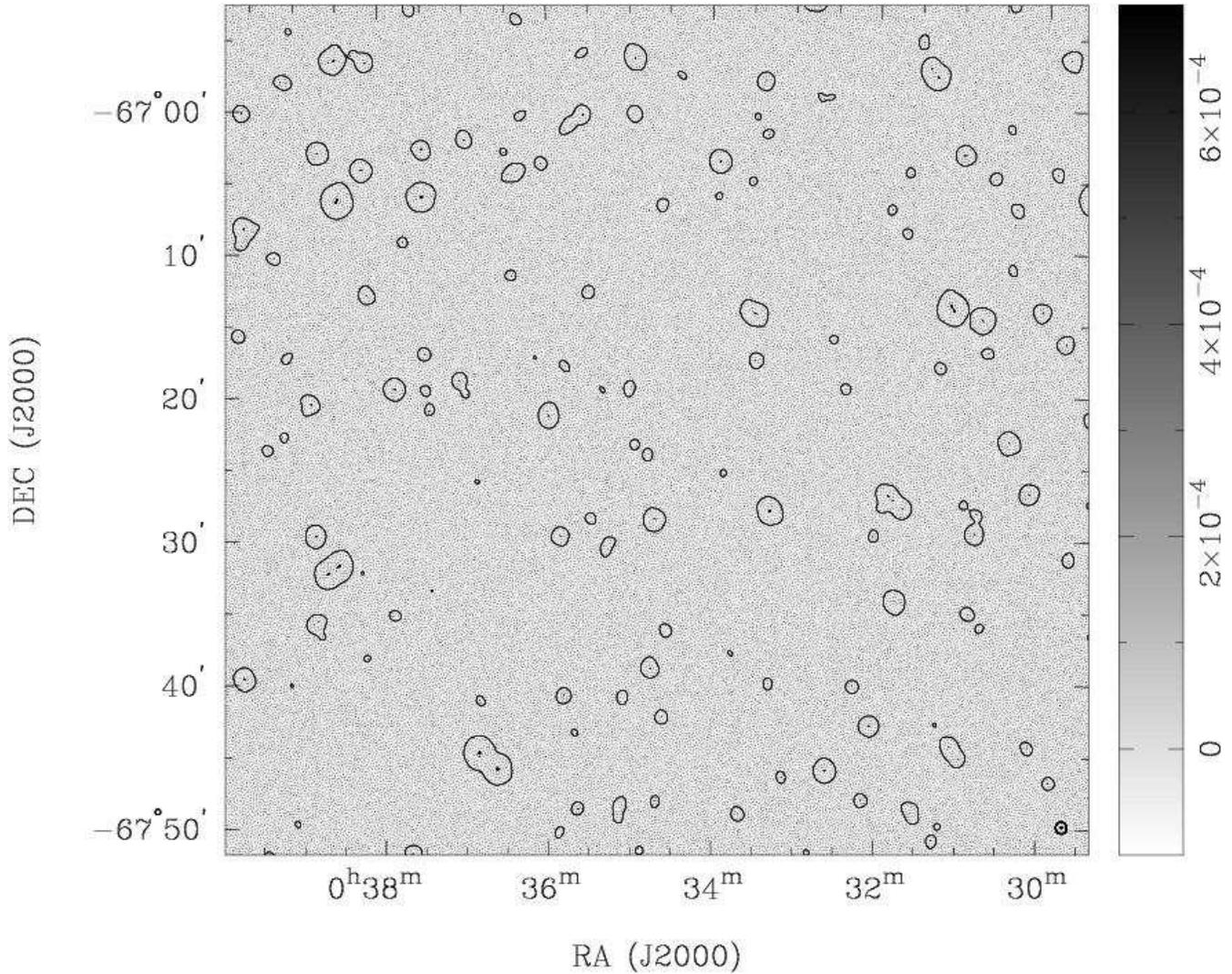}
\caption{A portion of ATLBS survey region A with the low resolution image represented
  by a contour at 0.4~mJy~beam$^{-1}$ and the high resolution image shown
  using grey scales.  The beam of the low resolution image has FWHM $52\farcs4
  \times 47\farcs4$ at P.A. $6\degr$ and the beam corresponding to the high
  resolution image is $4\farcs7 \times 4\farcs5$ at P.A. $-7\fdg8$; these
  beams are shown at the bottom right of the image as an unfilled ellipse and
  filled ellipse respectively.}
\end{figure*}

As we have noted above, the 750~m array configurations have, additionally,
sparse {\it uv}-coverage in the 3--5~km range as baselines to the sixth
antenna ca06. While we have taken every effort to ensure that the low resolution survey is
of the highest quality, we may examine the additional information provided by
the baselines to ca06 that could provide some critical information on the
compact components in the sources even though the {\it uv}-coverage is incomplete.
We have used these longer baselines
to construct independent images of the two survey regions with higher angular
resolution.  However, owing to the significant `hole' in the {\it uv}-coverage
below 3~km in this visibility dataset, the synthesized beam has significant
sidelobes and confusion is a serious issue.  Therefore, we have aided the
deconvolution by using the low-resolution images constructed using the 750-m
arrays to identify the sky regions that potentially contain source
components.  

The first step involved constructing a model representing sources detected
outside the primary beam main lobe using all the 750~m array data, including
the baselines to the 6-km antenna, and subtracting this from the visibility data.
The visibilities were then self-calibrated in phase, using models for source
components that were derived by imaging the primary beam main lobe region, and
the gains of the data 
in the two frequency bands were allowed a self-calibration
adjustment. The high resolution images were constructed using exclusively the
baselines to ca06, which correspond to a {\it uv}-coverage that sparsely fills the
annulus between 3 and 5~km. 
Deconvolution of these images was performed iteratively wherein the search region
for source components was progressively widened and the component search regions
in successive deconvolution iteration steps was derived from deconvolved images of
previous steps. In the final deconvolution step the search areas conformed to the regions
in the low resolution images---which were constructed using the 750-m 
baselines---in which the pixel intensity exceeded 0.4~mJy~beam$^{-1}$ (about $5~\times$ rms
noise). 

The high resolution images have been constructed using 5 instantaneous 
baselines per configuration compared
with 10 baselines used to make the low resolution images.  The rms noise is,
therefore, expected to be about 120~$\mu$Jy~beam$^{-1}$, and this expectation
is very close to that
measured in regions of the images that are apparantly source free.
Deconvolution iterations were stopped when the peaks in the residual image, within
the regions being searched for components, reduced below
0.4~mJy~beam$^{-1}$. This implies that in the high resolution images the
fractional flux density exceeding 0.4~mJy are deconvolved and restored with
Gaussian components; however, fractional intensities below 0.4~mJy continue to
be represented by beams with significant sidelobe structure and their integral
flux density over the image will be zero.

The size of the restoring beam following deconvolution was
determined by fitting Gaussian models to the main lobes of the synthesized
beams. The high resolution images of ATLBS survey regions A and B were made with beam
FWHM $4\farcs7 \times 4\farcs5$ at P.A. $-7\fdg8$ and $4\farcs8 \times 4\farcs4$ at
P.A. $3\fdg3$ respectively. Since the {\it uv}-coverage is an annulus, we might
expect that extended source components exceeding the size of these beams would be
resolved and their flux densities would be severely attenuated in these
images. Nevertheless, source components with size less than these beams would be
represented. 

As an example, a portion of survey region A is shown in Fig.~3, where
the high resolution image is in grey scales and the corresponding low
resolution image is represented using contours. Most
of the sources represented by closed contours also have compact components in
the grey scale image; sources that appear extended in the low resolution image
often appear to have multiple compact components.

\section{Properties of the radio sources in the survey} 

The radio images were scanned, sources identified, and their radio properties
listed by an automated algorithm.  The routine IMSAD in MIRIAD was modified
for this purpose. Only the sky region where the primary beam response in the mosaic exceeds
50\% was searched for sources, and flux density estimates were corrected for
the primary beam attenuation. 

Islands in the low resolution images, with connected pixels
exceeding 0.4~mJy~beam$^{-1}$, were considered to be independent
sources. Thus, the ATLBS source catalogue presented herein includes all
sources that have a peak flux density exceeding 0.4~mJy~beam$^{-1}$ in the low
resolution images. The
centroid position of these connected pixels was listed as the source position
and the source name was derived from this centroid position.  Sources were
automatically classified as unresolved, extended Gaussian or composite
based on their structure in the low resolution image and the success in modeling the
images using single-component Gaussians.  Peak and total flux
densities were estimated for each source, both in the low resolution image and
separately in the high resolution image. Extended sources without composite
structure were fitted using single Gaussian models and the fit parameters as well as
deconvolved source sizes were derived. 

The fractional integrated polarized intensity $\Pi_{0}$ was estimated for the ATLBS
sources from the low resolution images in Stokes Q, U and I.  Image pixels in
which the Stokes I intensity exceeded 0.4~mJy~beam$^{-1}$ were considered,
pixel intensities in Stokes Q, U and I were summed separately, a measure of
the integrated polarized intensity was estimated 
by computing $p_{m}=\sqrt{Q^2+U^2}$, where $Q$
and $U$ represent the pixel-summed image values in Stokes Q and U
respectively, the integrated polarized intensity was set to zero if the signal-to-noise
ratio in this estimate was less than unity, the polarized intensity estimate
was debiased (as described below) and
the fractional integrated polarized intensity $\Pi_{0}$ was computed as the
ratio of integrated polarized intensity to integrated total intensity. 

The measured integrated
polarized intensity $p_{m}$ was debiased with a simple first-order correction
to derive an estimate $p_{e} = \sqrt{p_{m}^2 - f_{p}\sigma_{p}^2}$, where
$\sigma_{p}$ is the standard deviation of the errors in Stokes Q and U image pixels
and $f_{p}$ represents the fractional increase in noise variance in the
pixel-summed values.  To estimate the fractional increase $f_{p}$, we assumed
(i) a Gaussian profile for the beam and, therefore, a Gaussian power spectrum for
the noise variance, and (ii) a `top-hat' function for the integration area
and, therefore, a $J_{1}(u)/u$ form for the noise filter function
corresponding to the pixel summation; $J_{1}(u)$ is the Bessel function of the
first kind.  The pixel averaging in the image domain is, in effect, a
convolution in the image domain that corresponds to a multiplication of the power
spectrum in the transform domain by a $J_{1}(u)/u$ form window.  Consequently,
the noise in the pixel-summed values would have a variance that is the
integral of this windowed power spectrum.
The noise variance in the pixel summed values $p_{m}^2$ is a
factor $f_{p}$ greater than the noise variance in the individual image pixels,
and $f_{p}$ is proportional to
\begin{equation}
 f_{p} \propto \int_{0}^{\infty} 2 \pi u {{J_1^2(2 \pi \theta_{\pi}
   u)}\over{(u / \theta_{\pi})^2}} exp \big[ - {{(\pi \theta_b)^2}\over{[-2
   ln(0.5)]}} u^2 \big] du,
\end{equation}
where $\pi (\theta_{\pi} / 2)^2$ is the summation area (in units of radians$^2$) and
$\theta_b$ is the FWHM of the beam (in units of radians).  The fractional increase $f_p$
versus the number of pixels in the summation is shown in Fig.~4.  
The ATLBS images have 1.75-arcsec pixels and, as expected, the variance in the
pixel summation rises as the square of the number of pixels
in the regime where the summation is over an area less than the beam area,
where as the variance in the pixel summation rises proportional to the number
of pixels in the regime where the summation area exceeds the beam area. The
break at about where the summation is over a beam area is because the noise in
image pixels is correlated within beam areas and uncorrelated on larger
scales.

\begin{figure}
\includegraphics[angle=-90,width=84mm]{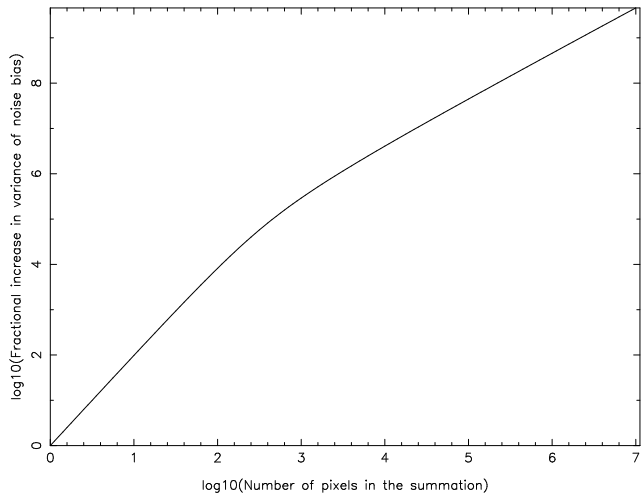}
\caption{The fractional increase $f_{p}$ in noise variance in pixel-summed image
  intensities is plotted as a function of the number of pixels in the
  summation. The computation is for ATLBS images with beam FWHM of $50\arcsec$
  and pixel size $1\farcs75$; there are about $10^{2.81}$ pixels within a beam
FWHM area and this is about where the curve has a break.}  
\end{figure}

In the case of extended sources the polarization position angle may vary over
the source; therefore, the integration of Stokes Q and U values over sources,
which correspond to measurements of Stokes parameters using beams in which the
sources are unresolved, may result in low fractional polarization.  The
fractional integrated polarized intensities $\Pi_{0}$ estimated above treat all 
sources over the entire range of flux densities as unresolved and these values
may be a useful comparison of fractional integrated polarizations versus flux density.


In the case of the imaging using the 750-m array data, the synthesized beam
has small amplitude sidelobes because the {\it uv}-coverage is almost complete.
However, in the case of the high resolution imaging the synthesized beam is
very different from a Gaussian approximation to the main lobe; therefore,
estimating integral flux densities of sources in the high resolution image
suffers uncertainties and requires careful understanding of the inherent
limitations arising from the annular visibility coverage and the cutoff level
adopted during deconvolution.  

Within the search region, the farthest distance of a source from pointing centres is
$16\farcm4$. Therefore, owing to the finite
bandwidth of the frequency channels in the multi-frequency continuum
data, visibility amplitudes at the longest baseline of 5020~m would 
be expected to have a worst case attenuation to 0.45 of the true value.
Simulations using the visibility coverage used to construct the high
resolution images and taking into consideration
the channel bandwidth corresponding to the observations suggest a
worst case attenuation of source peaks to 0.8 of their true value in the high
resolution images.  We have
modeled the bandwidth-related attenuation using a functional form fitted to
the simulation results and used this to scale the peaks derived from the high
resolution images.

In the high-resolution image, the search for the peak flux density is
over a `footprint' on the sky corresponding to the area enclosed by
the contour at 0.4~mJy~beam$^{-1}$ in the low-resolution image, or the contour
at half the peak flux density if this is lower. The median
size of the footprint, for sources with total flux density exceeding 0.4~mJy
is 107 beam areas (of the high-resolution image).  
Our algorithm estimates peak flux densities in the high
resolution images by searching for the peak within the corresponding
footprint.  The probability of a
spurious peak of flux density 0.4~mJy occuring in the high resolution images
and within the footprint area is below 5\%; therefore we may consider those
tabulated peaks exceeding 0.4~mJy~beam$^{-1}$ as reliable at the 95\% confidence level.

The integrated flux density of the sources in the high resolution image were
estimated by summing the flux densities in the compact components. An
iterative algorithm was adopted wherein the peak within the `footprint' was
located, a Gaussian fit made to the component corresponding to the peak and
the Gaussian component subtracted from the high resolution image. Successive
peaks---including positive and negative peaks---were located, fitted with
Gaussian components and subtracted until the 
absolute value of the peak residual was below a threshold of
0.5~mJy~beam$^{-1}$.  Thus, the listed values of integrated flux densities
in the high resolution images is an estimate of the total flux
density in compact components exceeding 0.5~mJy; in case the peak within the
source 'footprint' is less than 0.5~mJy then the listed integrated flux
density is simply the peak flux density.     

A total of 511 sources were identified in
Field~A and 585 sources in Field~B. The two fields have a slight overlap in
which there are two common sources; therefore, the number of sources in our
catalogue exceeding 0.4~mJy~beam$^{-1}$ peak flux density is 1094 over the
8.42 deg$^{2}$ sky area.

Of the 484 sources with total flux density in the range 0.4--1.0~mJy, 409
were classified as unresolved in the low resolution images made with beam FWHM
of $50\arcsec$.  Of the 75 extended sources (15\% of the sources in the
0.4--1.0~mJy range), 64 were deemed to
be representable using a single Gaussian model; only 11 sources were
classified as composite.  The fraction of sources classified to be extended
increases to 28\% in sources with flux density in the range 1--10~mJy and
about half of these sources are deemed composite in structure.  As much as
70\% of sources in the 10-100~mJy range in flux density are extended, with
three-fourths of these classified as composite. The fraction of sources that
are deemed extended and with structure well fit by single Gaussian models is
about 15\% over the entire flux density range 0.4--100~mJy; however, the fraction
that is deemed as possessing composite structure increases with flux density
rising from about 2\% for the sub-mJy population to above 50\% for sources
with 100~mJy flux density. 

The angular size for the ATLBS sources were estimated by measuring the area
enclosed by the contour at 0.4~mJy~beam$^{-1}$, assuming that the source has a
Gaussian profile, and computing the source width at half maximum.  This
estimate is expected to be conservative, and yields median angular size of
$10\arcsec$ for sources with flux density below 10~mJy, rising to
$20\arcsec$ for sources in the 10-100~mJy bin. As compared to the linear size
distribution derived by \citet{Wi90}, the median angular size of the ATLBS
sources appears substantially larger and
the distribution in angular size appears to cutoff more sharply.
Most ATLBS sources appear only 
marginally resolved in the low-surface-brightness images made with beam FWHM
of about $50\arcsec$, and we postpone a discussion of the angular size
distribution of the ATLBS sources to later papers where detailed
structural information is presented. 

A listing of the ATLBS radio sources is in Table~2, where we list the source
names and centroid positions, classification type using codes `P' for
unresolved sources, `G' for resolved sources that may be well 
fit with single Gaussians and `C' for sources with composite structure.
Total and peak flux densities derived from the ATLBS survey images of
$50\arcsec$ beam as well as those derived from the high resolution images are
listed.  Percentage integrated polarization $\Pi_{0}$ is in the last
column of the Table.

The absolute flux
density scale is based on the adopted flux density of the primary calibrator
PKS~B1934$-$638, which is within 2\% of the \citet{baa77} scale.
The absolute position of the phase calibrator has been measured with respect
to the International Celestial Reference Frame (ICRF) using long baseline
interferometers \citep{Ma98}.
The calibrated visibilities are expected to have 8\% amplitude errors because
of discrete source confusion within the primary beam during the calibrator
observations.  As a consequence, the 
initial model images, which serve as an input to the self-calibration and also
determine the absolute astrometry might have a systematic position error of $0\farcs1$.
Image noise contributes an rms error of (2/S$_{\rm mJy}$) arcsec in
positions of sources, which adds in quadrature to the systematic error
term (S$_{\rm mJy}$ is the source flux density in mJy). At the flux limit of
0.4~mJy, the position error may be as large as $5\arcsec$ or a tenth of the
synthesized beam FWHM. 
The image dynamic range is 1000, which implies that subsequent to
self-calibration we may have residual antenna based phase calibration errors of $0\fdg8$
rms or, equivalently, 1.4\% amplitude errors in the calibrated
visibilities used in the image making.  The errors in
peak and total flux density estimates are dominated by the image noise for sources below
about 4~mJy, and may be as much as 20\% for the faintest sub-mJy sources detected in
the survey. The listed total and peak flux densities were estimated from the
ATLBS low-resolution survey images and the corresponding high resolution
images; these images have rms noise of 0.085 and 0.12~mJy~beam$^{-1}$ respectively. 

\onecolumn
\begin{longtable}{@{}lllcrrrrr}
\caption{ATLBS sources. This table is presented in its entirety in the
  electronic edition of Monthly Notices at
  http://www.blackwellpublishing.com/products/journals/suppmat/MNR. A portion
  is shown here as a sample of ATLBS sources and their properties.}\\
\hline \\[-2ex]
Source name & RA & DEC & Type\footnote{Source types: P denotes an unresolved
  object, G denotes a single Gaussian 
component, C denotes a composite source.  $S_{\rm tot}$ and $S_{\rm peak}$ are
the total flux density and peak flux density of the source as measured using
the ATLBS low resolution images with $50\arcsec$ beam.  $S_{\rm int}^{\rm HR}$ 
and $S_{\rm peak}^{\rm HR}$ are, respectively, the integrated flux density of compact
components in the source and the peak flux density as measured using the high
resolution image of the source made with $4\farcs6$ beam.  The last column
lists $\Pi_{0}$, which is the percentage integrated polarization in the
source.} & $S_{\rm tot}$ & $S_{\rm peak}$ & $S_{\rm int}^{\rm HR}$ & 
$S_{\rm peak}^{\rm HR}$ & \% Pol. \\[0.5ex]
 & \multicolumn{2}{c}{(J2000.0 epoch)} & & (mJy) & (mJy) & (mJy) & (mJy) & $\Pi_{0}$
 \\[0.5ex]
\hline \\[-1.8ex]

J0038.9$-$6806	&	00:38:57.60	&	$-$68:06:03.6	&	G	&	1.45	&	0.96	&	0.79	&	0.79	&	20.2	\\ 
J0032.6$-$6805	&	00:32:39.62	&	$-$68:05:23.7	&	G	&	20.35	&	19.93	&	8.09	&	5.75	&	0	\\ 
J0035.7$-$6805	&	00:35:46.93	&	$-$68:05:45.8	&	P	&	0.87	&	1.05	&	0.97	&	0.97	&	37.3	\\ 
J0032.2$-$6804	&	00:32:17.02	&	$-$68:04:00.0	&	P	&	3.96	&	3.98	&	3.35	&	3.28	&	0	\\ 
J0033.0$-$6803	&	00:33:01.15	&	$-$68:03:33.1	&	P	&	0.97	&	0.86	&	0.95	&	0.95	&	0	\\ 
J0039.4$-$6801	&	00:39:24.57	&	$-$68:01:34.9	&	P	&	1.85	&	1.55	&	1.06	&	1.02	&	9.9	\\ 
J0034.1$-$6801	&	00:34:07.57	&	$-$68:01:45.4	&	P	&	1.97	&	1.95	&	1.06	&	1	&	6.9	\\ 
J0039.6$-$6800	&	00:39:41.03	&	$-$68:00:04.5	&	P	&	13.28	&	13.42	&	13.21	&	6.5	&	1.9	\\ 
J0039.1$-$6800	&	00:39:10.50	&	$-$68:00:35.4	&	P	&	0.59	&	0.61	&	0.57	&	0.57	&	0	\\ 
J0040.6$-$6800	&	00:40:36.30	&	$-$68:00:13.8	&	G	&	0.89	&	0.6	&	0.5	&	0.5	&	2.8	\\ 
J0037.5$-$6800	&	00:37:32.00	&	$-$68:00:24.5	&	P	&	0.88	&	1.05	&	1.09	&	1.07	&	17.6	\\ 
J0029.4$-$6759	&	00:29:25.83	&	$-$67:59:54.1	&	P	&	1.46	&	1.5	&	1.36	&	1.34	&	7.8	\\ 
J0036.2$-$6800	&	00:36:15.60	&	$-$68:00:18.5	&	P	&	0.78	&	0.87	&	1.14	&	0.96	&	0	\\ 
J0039.9$-$6759	&	00:39:57.84	&	$-$67:59:23.8	&	P	&	0.73	&	0.5	&	0.47	&	0.47	&	37.6	\\ 
J0035.5$-$6758	&	00:35:35.71	&	$-$67:58:48.7	&	P	&	1.7	&	1.64	&	1.93	&	1.68	&	0	\\ 
J0028.8$-$6758	&	00:28:51.34	&	$-$67:58:28.3	&	P	&	1.16	&	0.68	&	0.69	&	0.69	&	14.5	\\ 
J0030.8$-$6758	&	00:30:49.93	&	$-$67:58:11.2	&	G	&	1.11	&	0.95	&	1.11	&	0.91	&	9.2	\\ 
J0032.6$-$6757	&	00:32:39.23	&	$-$67:57:59.5	&	G	&	1.04	&	0.85	&	0.65	&	0.65	&	23.6	\\ 
J0038.1$-$6757	&	00:38:09.76	&	$-$67:57:09.2	&	P	&	1.39	&	1.32	&	1.37	&	1.19	&	0	\\ 
J0029.0$-$6755	&	00:29:00.17	&	$-$67:55:50.2	&	P	&	124.65	&	123.4	&	127.58	&	90.22	&	2.9	\\ 
J0033.6$-$6756	&	00:33:39.33	&	$-$67:56:37.0	&	P	&	1.75	&	1.68	&	0.55	&	0.55	&	17.6	\\ 
J0041.7$-$6756	&	00:41:45.48	&	$-$67:56:00.9	&	P	&	0.79	&	0.76	&	0.81	&	0.81	&	58.7	\\ 
J0027.2$-$6755	&	00:27:13.05	&	$-$67:55:01.3	&	P	&	5.4	&	5.32	&	1.98	&	1.74	&	5.2	\\ 
J0042.2$-$6755	&	00:42:16.02	&	$-$67:55:10.0	&	P	&	1.97	&	1.83	&	1.2	&	1.03	&	10.6	\\ 
J0040.4$-$6755	&	00:40:24.63	&	$-$67:55:12.2	&	P	&	0.99	&	0.88	&	0.8	&	0.71	&	4.2	\\ 
J0042.7$-$6754	&	00:42:43.52	&	$-$67:54:26.2	&	G	&	5.36	&	4.26	&	5.4	&	3.63	&	0	\\ 
J0035.6$-$6754	&	00:35:39.81	&	$-$67:54:48.0	&	P	&	0.77	&	0.63	&	0.63	&	0.53	&	12	\\ 
J0035.4$-$6754	&	00:35:24.59	&	$-$67:54:11.8	&	P	&	1.11	&	0.96	&	0.4	&	0.4	&	11.5	\\ 
J0028.9$-$6753	&	00:28:55.76	&	$-$67:53:34.6	&	G	&	1.59	&	1.42	&	0.51	&	0.51	&	8.6	\\ 
J0027.3$-$6753	&	00:27:18.51	&	$-$67:53:10.4	&	P	&	4.41	&	4.75	&	1.78	&	1.41	&	4	\\ 
J0034.2$-$6753	&	00:34:12.81	&	$-$67:53:22.9	&	C	&	2.44	&	1.66	&	0.63	&	0.54	&	5.4	\\ 
J0030.2$-$6753	&	00:30:15.01	&	$-$67:53:27.1	&	P	&	2.53	&	2.65	&	2.87	&	2.26	&	0	\\ 
J0039.5$-$6753	&	00:39:34.19	&	$-$67:53:37.6	&	P	&	0.62	&	0.57	&	0.45	&	0.45	&	0	\\ 
J0032.3$-$6753	&	00:32:20.55	&	$-$67:53:35.9	&	P	&	0.84	&	0.97	&	1.26	&	1.09	&	28.7	\\ 
J0036.0$-$6753	&	00:36:00.14	&	$-$67:53:25.2	&	P	&	0.41	&	0.51	&	0.51	&	0.51	&	58.4	\\ 
J0042.1$-$6752	&	00:42:08.14	&	$-$67:52:07.5	&	P	&	4.44	&	4.3	&	3.13	&	1.75	&	0	\\ 
J0033.1$-$6753	&	00:33:09.17	&	$-$67:53:26.1	&	P	&	0.41	&	0.44	&	0.4	&	0.4	&	32	\\ 
J0039.6$-$6752	&	00:39:40.53	&	$-$67:52:40.2	&	G	&	1.68	&	0.72	&	0.83	&	0.68	&	19	\\ 
J0039.4$-$6752	&	00:39:26.68	&	$-$67:52:12.2	&	P	&	1.25	&	1.28	&	1.51	&	1.29	&	15.1	\\ 
J0028.9$-$6751	&	00:28:54.12	&	$-$67:51:27.5	&	G	&	2.65	&	2.29	&	0.46	&	0.46	&	4.6	\\ 
J0033.3$-$6752	&	00:33:22.12	&	$-$67:52:19.1	&	P	&	0.46	&	0.58	&	0.45	&	0.45	&	21.5	\\ 
J0037.6$-$6751	&	00:37:39.87	&	$-$67:51:50.5	&	P	&	1.57	&	1.61	&	1.54	&	1.04	&	7.7	\\ 
J0041.2$-$6751	&	00:41:12.15	&	$-$67:51:23.6	&	P	&	0.66	&	0.55	&	0.48	&	0.48	&	0	\\ 
J0032.8$-$6751	&	00:32:49.14	&	$-$67:51:51.0	&	P	&	0.54	&	0.49	&	0.48	&	0.48	&	23	\\ 
\hline \\
\end{longtable}

\twocolumn
\subsection{Completeness and reliability of the survey}
 
Reliability of source detection is a major issue in most surveys, particularly
interferometer surveys that have poor visibility coverage. The ATLBS survey is
unique in that the entire survey regions are observed with
complete visibility coverage! Therefore, the synthesized beams are well
behaved and the reliability of the survey is determined by the
image thermal noise. The noise in the ATLBS images are well defined from the
data using Stokes V images: the rms noise is 85~$\mu$Jy~beam$^{-1}$.  
We have limited the catalogue to sources detected with a peak flux density
exceeding 0.4~mJy~beam$^{-1}$, which is 4.7 times the rms
noise.  At this level, the Stokes~V images have 4 peaks exceeding
0.4~mJy~beam$^{-1}$ over the entire survey area 
and these are in the range 0.4--0.5~mJy~beam$^{-1}$,
indicating that about 0.4\% of the sources in the catalogue might be spurious noise peaks and
that these spurious sources would be close to the flux limit of the catalogue.

In any survey image, the observed flux density of a source would be the true
flux density plus thermal noise.  Depending on the value of the noise at the
position of the source, the estimated flux density would be altered. When
sources are binned in flux density, noise results in movement of sources up or
down bins.  In effect, the source counts are
smoothed by a function whose width depends on the thermal noise in the image.
In the mJy and sub-mJy regime that is being explored in the ATLBS
survey, the differential source counts steeply decrease with increasing flux
density and, therefore, we expect that a net excess of faint sources would be detected
because some sources below the flux density cutoff would be noise
biased to lie above the detection threshold. 

To estimate the completeness and reliability of the ATLBS survey and assess
the effect of noise bias on the detection of sources, we have made simulations
in which sources were assumed to have a distribution in flux density
corresponding to the counts derived by \citet{Ho03}.  Owing to the image rms
noise of 85~$\mu$Jy~beam$^{-1}$, the expectation from the simulations is that the number of
sources detected in the 0.4--0.8~mJy bin would be enhanced by 16\%, that in
the 0.8--1.6~mJy bin would be enhanced by 2.5\% and that in octave bins at higher
flux density would be altered by less than a percent. As expected, the effect is greatest
close to the flux density cutoff.  18\% of the sources in the 0.4-0.8~mJy bin
are translated to adjacent bins---13\% to below 0.4~mJy and 5\% to above
0.8~mJy---but this is overcompensated by sources that are noise biased and translated
up in flux density from below the cutoff.  

To summarize, spurious sources are extremely rare and constitute 
only about 0.4\% of the
sources detected above the cutoff; 13\% of sources with true flux density in
an octave bin above the
cutoff would be noise biased to values below the cutoff and, therefore, would
fail to enter the catalogue; however, as much as 27\% of the sources detected
in the lowest octave bin of 0.4-0.8~mJy are expected to be genuine
sources with true flux density below the cutoff that are 
noise biased to lie above the detection threshold and, therefore, enter the catalogue.

\subsection{Confusion and source blending in the survey}

A limitation to the reliable detection of discrete radio sources in radio
surveys is confusion, which is because the radio image is a convolution of the
true sky with the telescope beam.  In surveys that are made using interferometer arrays
and with sparse visibility coverage, the synthesized beams have significant
sidelobes.  Along with the limitations to the dynamic
range arising from calibration errors, this makes it difficult to distinguish
all of the radio sources in the survey area although they may be above the
detection threshold as defined by the image thermal noise.  The confusion
limit on source detection is related to the filling factor in the visibility
plane.

The identification of discrete sources in the ATLBS survey was made using the
low resolution images, which were made with complete visibility coverage and,
therefore, with well defined synthesized beams. The source catalogue that is
the basis of the study of the radio source population was also derived from
this low resolution image. Therefore, classical confusion and its effects on source
selection and completeness and reliability are simply related to the finite
resolution of these ATLBS survey images.  A measure of the degree to which
confusion and source blending result in errors in source counts is how
sparsely sources are observed to cover the sky, which depends on the number of
beam areas that are observed to be occupied by sources. In the ATLBS low
resolution images, the number of effective beam areas per source detected is about 50.

Because of the low angular resolution in the ATLBS survey, confusion manifests
in blending of discrete radio sources. We have estimated the `blending
correction' from simulations.  Sources were assumed to span
a range 0.1--409.6~mJy in flux density and Poisson random distributed over the
sky area.  A distribution in flux density corresponding to 
the source counts derived by \citet{Ho03} was adopted, which was based 
on the Phoenix deep survey (PDS).  Sources were deemed to be confused if
they were connected by a contour at a level of 0.4~mJy~beam$^{-1}$, which was
the criterion adopted while forming the ATLBS source catalogue. 
Confusion between sources in different
flux density bins, as well as blending of multiple sources, was allowed for in
the simulations.  The high surface density of sources with low
flux density and the large sky areas covered by the point spread functions
associated with sources with relatively higher flux density together result in a
confusion between weak sources with the stronger sources.  
The simulations revealed that for the adopted PDS source counts
and the image point spread function corresponding to the ATLBS, confusion results in a 
significant reduction in the number of sources in octave-band flux-density
bins below about 1~mJy, and a small fractional increment in source counts
at higher flux densities. Based on the simulations, we estimate that the
blending correction factor, owing to classical confusion, is as much as 1.2 in
a 0.4--0.8~mJy bin, about 1.02 in a 0.8--1.6~mJy bin, and less than 2\% in higher flux
density bins.

The high resolution images were constructed from sparse visibility coverage
and, consequently, confusion is indeed an
issue that limits the elucidation of the high-resolution structures.  Only 1\%
of the visibility plane is covered by the data used to reconstruct these images.
As discussed above, the high-resolution images were 
examined only in the regions where sources
were identified in the low resolution images.  Operationally, we have used the
low resolution images to define the source regions as a constraint during the
deconvolution of the high resolution images.  This restricted the sources in
the high resolution image to be within 2\% of the total survey area and
resolves the ambiguities arising from the poor visibility coverage and consequent 
confusion in the high resolution imaging.
We estimate that even if every one of the discrete sources identified in the
ATLBS survey were a triple, the number of beam areas per source component
would be 20 in the case of the high resolution image. Moreover, in this
manuscript we restrict to using only estimates of the peak and integrated flux
densities of the high resolution images to derive indicators for the source complexity and
diffuseness. Detailed high-resolution radio structures of the ATLBS sources,
based on followup observations that provide improved visibility coverage, will be
presented in later ATLBS related publications.

\subsection{The source counts}

The number density of ATLBS sources detected with peak flux density exceeding
0.4~mJy~beam$^{-1}$ is 130 sources per square degree.  

The normalized differential source counts derived from the source list are
shown in Fig.~5. The total flux densities were binned in octave bin
ranges: 0.4--0.8, 0.8--1.6, 1.6--3.2 and so on till 204.8--409.6~mJy. We plot the
normalized differential source counts versus the mean flux density $\langle S
\rangle$ of sources in the individual bins, the differential counts $dN/dS$ were 
normalized to $\langle S\rangle^{-2.5}$. 

The total flux densities of the sources were estimated
from the low resolution image: the Gaussian fit parameters were used to infer
the total flux densities in the case of unresolved sources as well as
those sources with simple structure that were well fit using a single Gaussian
model. Composite sources were identified by the significant deviation in the
image pixels from the best fit Gaussian model---most of these sources appeared
to be 
composed of multiple components---and in these cases the total flux
density was derived by summing the image pixels over the source area. 
During the synthesis imaging that made the low-resolution image, the iterative
deconvolution had been terminated at the 1-$\sigma$ noise level 
and as a result the point spread function for the residual noise slightly
differs from the restoring beam used to convolve the CLEAN components: thus
the effective beam for weak sources slightly differs from the restoring beam. 
The algorithm that does
the Gaussian fit was limited to using only image pixels exceeding 2-$\sigma$
and the fit was weighted by the image pixel intensity to ameliorate the error
in the estimation of source flux densities arising from this limitation in
image fidelity. Nevertheless, there
were a significant number of sources with peak flux density
exceeding 0.4~mJy~beam$^{-1}$---30 in all---in which the Gaussian fit parameters were
somewhat smaller than the beam size and the estimate of total
flux density for these relatively weak sources falls below 0.4~mJy. The source
counts in the lowest bin of 0.4--0.8~mJy might be underestimated by 10\% due
to this effect, and the error bar for this bin has been enhanced to reflect
this additional source of uncertainty.  

\begin{figure}
\includegraphics[width=84mm]{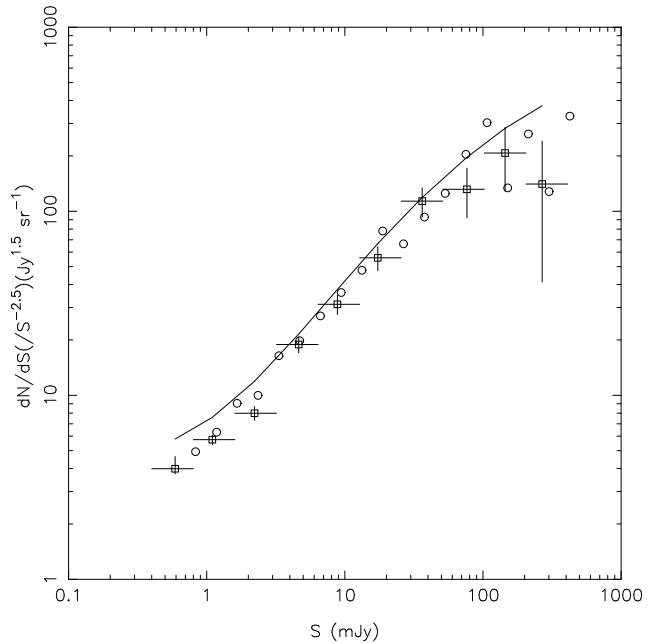}
\caption{Normalized differential source counts. The counts derived
  from the ATLBS 
  survey are shown using square box symbols; the horizontal bar associated
  with each symbol spans the bin range corresponding to the count and the
  vertical bar is the 1-$\sigma$ error bar. As a comparison, ATESP survey counts
  \citep{Pr01} are plotted using
  circle symbols and the fit to PDS counts \citep{Ho03} is shown as a
  continuous line.} 
\end{figure}

\begin{table}
\caption{ATLBS source counts}
\begin{tabular}{@{}rcrr}
\hline
$\Delta S$ & $\langle S \rangle$ & $N_{s}$ 
& $dN/dS(/S^{-2.5})$ \\
(mJy) & (mJy) & & (Jy$^{1.5}$~sr$^{-1}$) \\
\hline
0.4--0.8 & 0.59 & 364 & 3.99 ($+0.65,-0.25$) \\
0.8--1.6 & 1.10 & 289 & 5.74 ($\pm 0.35$) \\
1.6--3.2 & 2.22 & 140 & 7.99 ($\pm 0.68$) \\
3.2--6.4 & 4.63 & 104 & 18.9 ($\pm 1.9$) \\
6.4--12.8 & 8.83 & 70 & 31.2 ($\pm 3.7$) \\
12.8--25.6 & 17.3 & 46 & 55.8 ($\pm 8.2$) \\
25.6--51.2 & 36.3 & 30 & 113.5 ($\pm 20.7$) \\
51.2--102.4 & 76.5 & 11 & 131.6 ($\pm 39.7$) \\
102.4--204.8 & 144.5 & 7 & 207.2 ($\pm 78.3$) \\
204.8--409.6 & 268.4 & 2 &  140.7 ($\pm 99.5$) \\
\hline
\end{tabular}

\medskip
The first column lists the bin ranges for the total flux density, the second
column lists the mean flux density of the sources in the bins, the third
column lists the number of sources detected in each of the flux density
ranges and the last column lists the normalized differential source counts
corrected for various factors discussed in the text. 
\end{table}

The source counts in the flux density bins have been corrected for the primary
beam attenuation over the survey area by scaling the 
count corresponding to each detected source by the ratio of the total area of
the survey and the area over which the source is detectable: since we truncate
at the level where the primary beam drops to 50\%, this correction is relevant only
for the 
lowest octave bin and in this bin the counts were scaled up by a factor 1.33
to account for this effect. 

There may be sources with
peak flux density below 0.4~mJy~beam$^{-1}$, and missing in the derived source
catalogue, which are extended and have total flux density exceeding
0.4~mJy: the derived ATLBS source counts would be an underestimate due to such
sources as well. However, analysis of the source
structural properties in Sections~5.5 and 5.6 suggests that most sources have simple
structures at low flux density levels, at least at the resolution of the ATLBS
survey; therefore, we expect any correction owing to extended source structure
to be small. If we adopt the model for the angular size
distribution of sources derived by \citet{Wi90}, which has the median radio
source angular size $\Psi_{\rm med}$ of sources 
with 1.4~GHz flux density $S_{\rm 1.4}$ (in mJy) equal to 
$2.0 S_{1.4}^{0.30}$~arcsec, and an exponential form for 
the integral angular size distribution, the resolution corrections required to
be made to our derived source counts are less than 1\%. 

The ATLBS source counts shown in Fig.~5 have been  
corrected for the primary beam attenuation, noise bias that was discussed in
Section~5.1, 
resolution and blending; these corrected counts are also listed in Table~3.
The second column of the Table lists the mean flux density of sources in the
bins, column 3 lists the number of sources detected in the individual bins
(uncorrected) and the last column lists the derived normalized differential
source counts corrected for the effects discussed above.

As a comparison, ATESP survey counts \citep{Pr01} and the fit to PDS 
counts \citep{Ho03} are also shown in Fig.~5. The PDS counts were derived
from their survey that covered 4.56 deg$^{2}$ area and the ATESP survey
covered 26 deg$^{2}$ area; these are comparable to our survey area of 8.4 deg$^{2}$.

Within the errors, the observed counts appear consistent with
the estimates derived by \citet{Pr01} based on the ATESP survey.
The counts are, however, systematically lower than that estimated by
\citet{Ho03} based on the PDS: in the flux density range 0.8--200~mJy our ATLBS
counts are on the average a factor 0.8 of the PDS counts.  
As was the case for the ATESP survey counts, which do not 
show an upturn and suggest that any upturn at faint flux density levels is
below 1~mJy, the ATLBS counts are also consistent with no upturn down to about 0.6~mJy.

At the faint end of the flux
density scale explored by the ATLBS survey, blending reduces the
observed counts where as noise bias enhances counts. The
blending corrections, as well as our estimates for the enhancement in the
observed counts owing to noise bias, might be overestimates because we have assumed the 
relatively higher counts, based on the PDS survey, for the simulations that
estimate these correction factors.  Additionally, as discussed above, the counts in
the lowest bin might be an underestimate due to missing sources because their
peaks may be below the 0.4~mJy~beam$^{-1}$ cutoff or because of image errors
arising from the deconvolution algorithm adopted.

A cause for
significant discrepancies between different estimates of the source counts at
the sub-mJy levels might be field-to-field variations.  The
relatively large sky area, spread over multiple sky patches, covered by the
ATESP survey makes these counts a 
more reliable indicator of counts at these faint levels and there is good
agreement between the ATLBS counts and the ATESP counts in the 0.4--1.6~mJy
bins: both are systematically low compared to the PDS survey counts.

The systematic low counts derived from the ATLBS survey is
not owing to classical confusion; the ATLBS counts presented here have been corrected for
blending confusion. The low counts, relative to the PDS, could be interpreted as arising
due to a 30\% 
underestimate in the flux density of sources, or a 20\% reduction in numbers
of sources.  Since the ATLBS has good surface brightness sensitivity, it is
expected that the source catalogues derived from the survey would not have any
missing flux density.  Additionally, the ATLBS survey is potentially capable
of detecting extended sources with low surface brightness, which may be missed
in other surveys.   The low counts are more likely 
indicating that the ATLBS survey detects a smaller number of sources as
compared to surveys like the PDS.  It has been pointed out by \citet{Ho03}
that the PDS counts are based on a component catalogue, rather than a source
catalogue as was the case for the ATESP counts.  Source counts derived from
such component catalogues would be expected to overestimate the numbers of
sources as a result of sources in higher flux density bins
degenerating into multiple sources in bins with lower flux densities. The
ATLBS survey is a source catalogue owing to the relatively large size of the
synthesized beam; therefore, it is unsurprising that the counts derived are
consistent with that of the ATESP survey and below the PDS counts. 

\subsection{The fractional polarization in ATLBS sources}

As discussed above in Section~5, we have derived estimates for the 
percentage integrated polarization $\Pi_{0}$ for the ATLBS sources using the low
resolution images, integrating Stokes Q, U and I over
the source, computing the debiased polarised intensity and dividing by the
total intensity. The sources were 
binned in flux density, and the median integrated percentage polarization $\langle \Pi_{0}
\rangle$ as well as the
mean flux density $\langle S_{\rm mJy} \rangle$ of the sources in the individual
bins were computed; these 
values are in Table~4. The errors have been estimated using the Efron bootstrap method
\citep{ef79}, in which the samples in each bin were randomly
resampled with replacement to derive a sampling distribution of the median
and, thereby, an estimate of the error in the median.  
In Fig.~6 we plot the median integrated percentage polarization $\Pi_{0}$ versus
mean flux density for the binned data. The plot 
clearly shows an increasing fractional polarization
with decreasing flux density.  This trend is consistent with the
observation that among polarized radio sources, the faint sources are more
highly polarized than the relatively stronger sources \citep{tay07}.

The binned data were fitted to a power-law to derive the trend: 
\begin{equation}
\langle \Pi_{0} \rangle = 10.56 \langle S_{\rm mJy} \rangle ^{-0.565},
\end{equation}
where  $\langle \Pi_{0} \rangle$ is the median percentage integrated 
polarization and $\langle S_{\rm mJy} \rangle$
is the mean flux density (in mJy). The fit is also shown in Fig~6.
The data appears well fit by this single power-law form over more than a decade in
flux density; however, there appears to be a flattening above 10~mJy
suggesting that at higher flux densities the fractional integrated
polarization in sources may not change with flux density.

\begin{table}
\caption{ATLBS sources: percentage integrated polarization}
\begin{tabular}{@{}rrr}
\hline
{Flux density} & {Mean flux density} & {Median integrated } \\
{bin (mJy)} & {(mJy)} & {\% polarization $\Pi_{0}$} \\
\hline
0.4--0.8 & $0.59 \pm 0.006$ & $14.15 \pm 0.95$ \\
0.8--1.6 & $1.10 \pm 0.013$ & $10.5 \pm 0.66$ \\
1.6--3.2 & $2.22 \pm 0.039$ & $6.25 \pm 0.59$ \\
3.2--6.4 & $4.63 \pm 0.091$ & $4.25 \pm 0.46$ \\
6.4--12.8 & $8.83 \pm 0.21$ & $3.00 \pm 0.36$ \\
12.8--25.6 & $17.3 \pm 0.50$ & $2.50 \pm 0.45$ \\
25.6--51.2 & $36.3 \pm 1.28$ & $3.10 \pm 1.31$ \\
\hline
\end{tabular}

\medskip
The total flux density of sources has been used in the binning. The
quoted errors are 1 standard deviation (1-$\sigma$) values. 
\end{table}

\begin{figure}
\includegraphics[width=84mm]{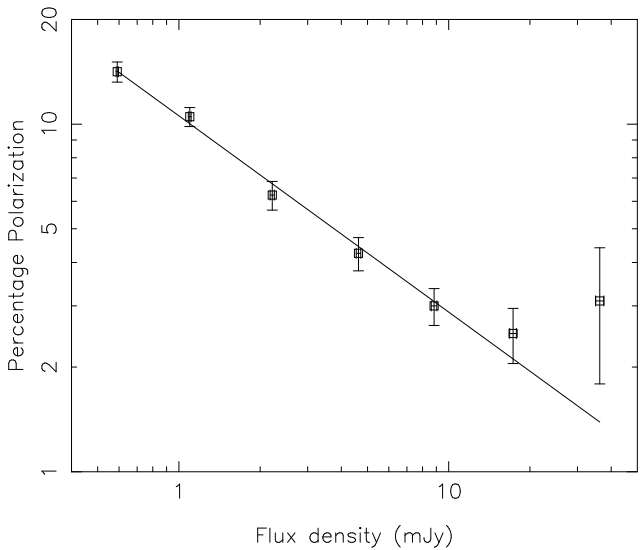}
\caption{ Median integrated percentage polarization $\Pi_{0}$ 
  versus flux density. The sources were
  binned in flux density, and the median percentage integrated polarization as well as
  the mean flux density of the sources in the individual bins were
  computed. Error bars are $\pm 1$ standard deviation. A single power-law fit to the
  data is also displayed.}
\end{figure}



We have adopted a signal-to-noise ratio based cutoff that sets to zero
estimates of polarized intensity that are below one standard deviation of the
expected noise.  Such a cutoff may potentially result in a positive residual
polarization bias if the polarized intensity has a low signal-to-noise ratio
\citep{Lea89}. The estimator of fractional integrated polarization, $\Pi_{0}$,
was separately recomputed assuming a signal-to-noise ratio based cutoff that set to
zero estimations that are below two standard deviations of the expected noise:
in this case as well
the mean percentage polarizations continued to display the trend of increasing
fractional polarization with decreasing flux density, although the residual
polarization bias in this case is expected to be negative at low signal-to-noise ratios.
This test adds weight to our finding that the fractional polarization in radio
sources increases with decreasing flux density.



\subsection{Complexity of radio sources at sub-mJy flux density}

\begin{figure}
\includegraphics[width=84mm]{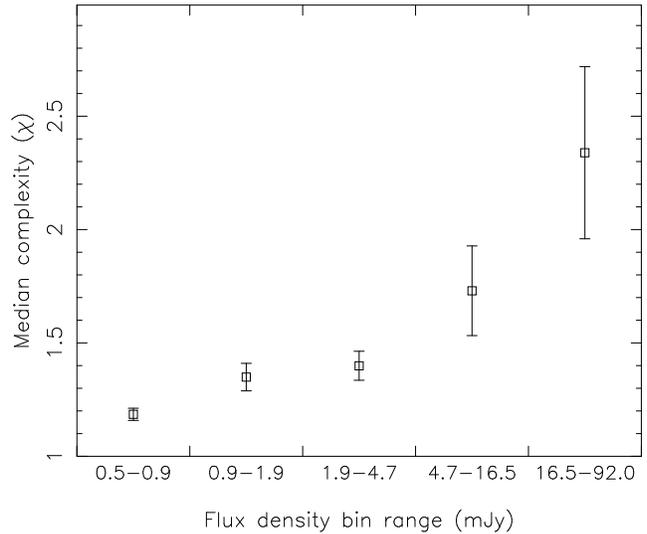}
\caption{Median complexity $\chi$ of sources in the ATLBS survey.  The sources
have been binned in ranges in flux density; the total flux
density estimated from the low resolution ATLBS survey images were used in the
binning. Error bars are $\pm 1$ standard deviation.}  
\end{figure}

\begin{table}
\caption{ATLBS source complexity}
\begin{tabular}{@{}rrrr}
\hline
{Flux density} & {Median flux} & {Number of} & {Median} \\ 
{bin (mJy)} & {density (mJy)} & {sources} & {complexity $\chi$} \\
\hline
0.5--0.9 & 0.67 & 339 & $1.19 \pm 0.03$ \\
0.9--1.9 & 1.21 & 271 & $1.35 \pm 0.06$ \\
1.9--4.7 & 2.72 & 148 & $1.40 \pm 0.06$ \\
4.7--16.5 & 7.18 & 143 & $1.73 \pm 0.20$ \\
16.5--92.0 & 27.85 & 63 & $2.34 \pm 0.38$ \\
\hline
\end{tabular}

\medskip
The total flux density of sources has been used in the 
  binning. The quoted errors for the median complexity are 1 standard
  deviation (1-$\sigma$) errors. 
\end{table}

The radio sources were identified in the low resolution images as a set of
connected pixels with intensity exceeding 0.4~mJy~beam$^{-1}$.  These images
have a high surface brightness sensitivity and were
made with a beam of FWHM about 50$\arcsec$.  The total flux densities were
estimated either from a Gaussian fit to the source image pixels in this low
resolution image or, in the case of sources with composite structure, from a
summation over the image pixel intensities.  Apart from deriving a value for
the peak flux density in the low resolution image, we have also separately derived the
peak flux densities in each of the sources from the high resolution images, which were made
with beam FWHM about $4\farcs6$. 

The ratio of the total flux density, as measured using the low resolution
image, to the peak flux density, as measured in the high resolution image, is
a measure of the departure of the source appearance from that of an unresolved
object of size well below $4\farcs6$. This ratio $\chi$, which we
adopt as a measure of how complex sources appear to be, is a measure of how complex the
source structure is when observed with a beam of FWHM $4\farcs6$. $\chi$ is
expected to be unity for unresolved sources, and exceed unity for resolved
sources. $\chi$ will exceed unity not only in the case of sources with
extended emission that is resolved by a $4\farcs6$ beam, but also in the case
where the source is composed of multiple components (which may be individually
unresolved). Additionally, $\chi$ will exceed unity in cases where the source
in the low resolution image is confused, for example, because two or more
unresolved and unrelated sources lie close together on the sky and within the
$50\arcsec$ beam of the low resolution image.  As discussed in Section~5.2,
such blending is not expected to be an issue at flux density exceeding 0.8~mJy,
and the effect of blending on source counts at these higher flux density
levels is expected to be less than 2\%.

There are 1063 sources in the ATLBS survey with total flux density exceeding
0.4~mJy. The median $\chi$ for these sources is 1.28. 20\% of the sources
have $\chi$ exceeding 2.0, which implies that when observed with beam FWHM
$4\farcs6$, close to a fifth of the ATLBS sources are either doubles, triples or
more complex sources or have more than half of their total flux density
in extended emission components. 

In Table~5 we list the median complexity of sources in bins of 
total flux density. 
In Fig.~7 is shown the distribution of median complexity $\chi$ 
versus total flux density. The errors
listed in the table, as well as the error bars in the figure, correspond to 1
standard deviation errors that were estimated from the data using the Efron
bootstrap method \citep{ef79}. The bins were chosen to have widths increasing
with flux density; bin widths are proportional to (flux density)$^{3/2}$. 
Fig.~7 shows that the source complexity
increases with increasing total flux 
density. This is consistent with earlier findings that the median angular size
of sources declines towards lower flux density, and that fainter radio sources
are increasingly compact.  

The derived source complexity $\chi$ has errors owing to the error in the estimates
for source total flux density and the peak flux density in the
high-resolution image.  Values of total flux density have rms errors less
than 20\%; errors are less than 10\% in sources with flux density exceeding
1~mJy.  The peak flux density in the 
high resolution image has an absolute rms error of 0.12~mJy. 
As discussed above in Section~5, the median source `footprint', which is the
search area for the 
peak in the high resolution image, has 107 (high resolution image) beam areas.
It follows that the probability of a chance peak within the `footprint'
exceeding 0.45~mJy is less than about 1\%. Spurious noise peaks within the
footprint, which exceed the true peak flux density of the source, would result
in an over estimate of the peak flux density and lead to an underestimate
for the source complexity.  The estimate for source complexity $\chi$ would,
therefore, be a lower limit to its true value and this underestimation would
be greater in sources with smaller flux density.

If we consider sources in the 0.5--0.9~mJy bin, which has the sources with 
lowest flux density, the median flux density is 0.67~mJy and the sources 
are estimated to have a median complexity of $1.19 \pm 0.03$. For sources in
this bin, the probability that a noise peak within the source footprint exceeds 0.45~mJy
and, consequently, the complexity is underestimated to have a value below 1.19 is
less than 1\%. Therefore, it is unlikely that image noise causes the median
complexity of sources in the 0.5-0.9~mJy bin to be as low as 1.19. The effects of
image noise are less significant in the source complexity estimates at higher
flux densities. The low value for the source complexity estimated for the
sub-mJy ATLBS population, and the rise in complexity with flux density, are
likely to be genuine.

If we consider sub-mJy sources in the flux density range
0.7--1.0~mJy, for which the flux densities all exceed six times the rms noise in
the high resolution image, a significant number of these sources may have
complexity exceeding 2.0 despite the noise.  Even if all of these sources have
true $\chi$ exceeding 2.0, only in about a sixth of these
sources do we expect the noise peaks in the high resolution image to result in
estimates for the complexities below 2.0.  We find that only 8\% of sources in
this flux density range have complexity exceeding 2.0. In contrast,
28\% sources in the flux density range 1--10~mJy have complexity exceeding
2.0, and 55\% sources in the 10-100~mJy range have complexity exceeding 2.0. 

\subsection { Diffuse emission associated with sub-mJy radio sources}

The parameter $\chi$ introduced above is a measure of the degree of source
complexity, but does not distinguish sources with extended or diffuse emission
from sources with multiple components, composite structure composed of compact
components, and confusion.

\begin{figure}
\includegraphics[width=84mm]{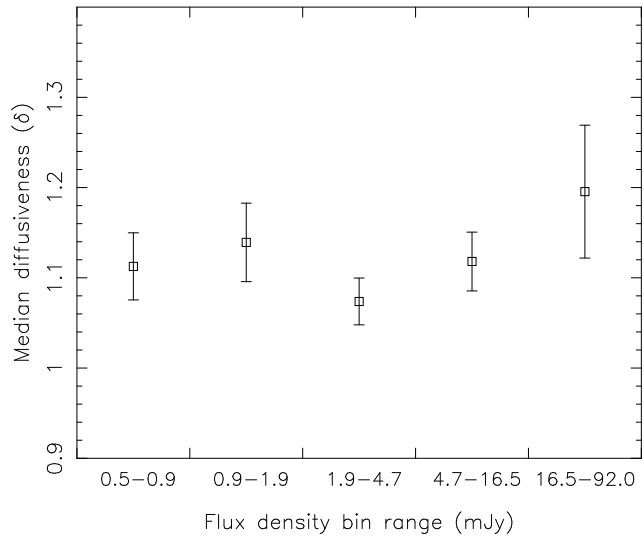}
\caption{Median value of the degree of diffuse emission $\chi$ of sources in
  the ATLBS survey.  The sources 
have been binned in ranges in flux density; the total flux
density estimated from the low resolution ATLBS survey images were used in the
binning. Error bars are $\pm 1$ standard deviation.}  
\end{figure}

\begin{table}
\caption{Degree of diffuse structure in ATLBS sources}
\begin{tabular}{@{}rrr}
\hline
{Flux density} & {Number of sources} & {Median diffuseness} \\
{(mJy)} &  & {$\delta$} \\
\hline
0.5--0.9 & 339 & $1.11 \pm 0.04$ \\
0.9--1.9 & 271 & $1.14 \pm 0.04$ \\
1.9--4.7 & 148 & $1.07 \pm 0.03$ \\
4.7--16.5 & 143 & $1.11 \pm 0.03$ \\
16.5--92.0 & 63 & $1.20 \pm 0.07$ \\
\hline
\end{tabular}

\medskip
The total flux density of sources has been used in the binning. The quoted
errors for the median diffuseness are 1 standard deviation (1-$\sigma$) errors.
\end{table}

The high resolution images described in Section~4 were constructed using a
visibility coverage that is an annulus and the beam FWHM is about $4\farcs6$;
therefore, it is expected that extended emission on scales exceeding
$4\farcs6$ would be resolved out and absent in these images. The fractional
flux density in extended emission, which would be missing in
the $4\farcs6$ resolution images, may be characterized by the ratio of the
total flux density in the low resolution images to the integrated flux
density in the high resolution images.  We refer to this ratio parameter as
$\delta$, representing the degree of diffuse emission in the source or, in
other words, a diffuseness parameter.  This parameter is expected to reveal
the quantum of flux density in extended diffuse emission.

The median value of $\delta$ for the sources with total flux density
exceeding 0.4~mJy in the ATLBS survey is $1.09 \pm 0.02$; about half of the
sources have more than a tenth of their flux density in diffuse emission.  
The median $\delta$
is significantly smaller than the median $\chi$, which is 1.28, as might be
expected since the complex structure is only partly diffuse structure. About 10\%
of the ATLBS sources with flux density exceeding 0.4~mJy~beam$^{-1}$ have $\delta$
exceeding 2.0; these sources have over half of their total flux density in diffuse
emission. 

We have computed the median value of the degree of diffuse structure---median
$\delta$---in bins of total flux density, where the source total flux density is
determined from the low resolution images with high
surface brightness sensitivity.  In Fig.~8 is shown the variation in $\delta$
with total flux density; the values are in Table~6.  The errors were estimated from
the data, as above, using the Efron bootstrap method. The ATLBS radio source
population does not show any significant trend in the 
degree of diffuse emission $\delta$ versus total flux density, although the source
complexity $\chi$ shows a significant 
rise towards higher total flux density.  The increased complexity in sources with
higher flux density appears to be owing to the sources 
being composed of multiple compact components rather than increased fraction
of diffuse emission. The fractional flux in diffuse emission appears to be 
fairly constant, independent of total flux density, at least in the range 0.4--100~mJy.

The total flux density is determined from the low resolution images with
rms noise 0.085~mJy~beam$^{-1}$.  The fractional error in this estimate is
at most 20\%, and less than 10\% for sources with total flux density exceeding 1~mJy.
The integrated flux density estimate derived from the high-resolution image is a
summation of the flux densities in compact components with peak exceeding
0.5~mJy; in those cases where no compact component exceeds this threshold the
estimate for integrated flux density is simply the value of the peak within
the `footprint'. Therefore, the tabulated values of
integrated flux density of compact components, which are used in estimating
$\delta$, represent a lower limit to the integrated flux in compact components
and may miss compact components that have peak flux density below
0.5~mJy~beam$^{-1}$ in the high resolution image. On the other hand, 
the high resolution images have an rms noise of 0.12~mJy~beam$^{-1}$ and in
cases where the source does not have compact components above this noise the
noise peak within the `footprint' would be tabulated as the integrated flux density of
compact components.  There is less than 1\% chance of a noise peak exceeding
0.45~mJy within the footprint; however, peaks exceeding 0.35~mJy are expected
with 25\% probability.  Weak compact components that are missed result in an
overestimate in $\delta$, noise in the high resolution image results in
underestimates for $\delta$.  
Owing to the image rms noise, sources with a given total flux density are
unlikely to have $\delta$ exceeding an upper bound; for example, it is
unlikely that sources with total flux density below 0.56~mJy have $\delta > 2.0$.

We define the sky area of any source to be the area enclosed by the
0.4~mJy~beam$^{-1}$ contour in the low resolution image of the source.
We have estimated the integrated flux density in the compact components in any
source by summing the flux densities of all the components in the high
resolution image that are located within the
sky area of the source.  This integrated flux density exceeds the peak flux density of the 
source (as measured in the high resolution image) in most cases, as expected.
The excess may be quantified as the ratio of the
integrated to the peak flux density; both measured in the high resolution images.
We find that this excess increases with increasing
flux density.  36\% of sources with flux density exceeding 10~mJy have this
ratio of the integrated to peak flux density exceeding 2 and 10\% of sources in
the 1--10~mJy range have 
integrated flux density exceeding the peak by a factor 2 or more. In the
sub-mJy population of sources that have an integrated flux density in the 0.4--1.0~mJy range
(484 sources), where estimates of this ratio may be considered to be lower
limits, less than 1\% of sources have listed integrated flux density exceeding
the peak by a factor of 2.  This finding is consistent with a change in the
radio structure of ATLBS sources with flux density in which the abundance of
multiple compact components is greater in sources with higher flux density:
the fainter sources may be dominated by sources with a single compact component
whereas brighter sources have double and triple compact structures.
 
\section{Summary}

We have used the Australia Telescope Compact Array to survey 8.42~deg$^{2}$ 
sky area at a radio frequency of 1388~MHz.  The interferometer observations were
made in a mode designed to mosaic image the wide field with complete
visibility coverage, and hence low confusion, and with exceptional
surface brightness sensitivity.  The data were used to reconstruct (a) a low
resolution image, with beam FWHM $53\arcsec \times 47\arcsec$ and rms noise
0.08~mJy~beam$^{-1}$, and (b) a high resolution image with beam FWHM of
$4\farcs6$. Whereas the low resolution image reproduces the extended and
diffuse radio emission associated with sources in the fields, the high
resoluion image resolves out structure on scales exceeding the beam
size. Together, the images provide an estimate of the structural properties of
the mJy and sub-mJy radio sources. We refer to our radio survey as the
Australia Telescope low brightness survey, and use the acronym ATLBS.

A total of 1094 radio sources with peak flux density exceeding
0.4~mJy~beam$^{-1}$ in the low resolution image were cataloged and their
source properties estimated. The source detections correspond to a density
of 130 sources per square degree. The studies presented herein have considered
only sources with peak flux density exceeding about 5 times the rms noise in
the image. 

The normalized differential source counts derived from the ATLBS
shows no evidence for an upturn down to about 0.6~mJy; the ATLBS counts are
consistent with the ATESP source counts, but relatively low compared to many
other surveys, including the PDS. This result suggests that there is no
substantial population of low surface 
brightness sources or source components at these flux densities that have been
missed by previous surveys.  The
ATLBS counts---as also ATESP counts---are relatively low compared to many
other counts perhaps because our counts are based on a source catalog,
rather than a component catalog.
As far as we know, blending has been considered and corrected for
the first time in the work presented herein; blending is of concern in surveys
such as the ATLBS that aim to go deep in surface brightness sensitivity. 
The derived ATLBS source counts have been corrected for blending, noise bias,
resolution and primary beam attenuation over the survey area.

The derived source counts are consistent with that of the ATESP survey suggesting
that the relatively large sky coverage of these surveys is key to robust
measurement of source counts. The considerable scatter in the derived
differential source counts at about 1~mJy flux density between the various
surveys that were made with relatively smaller sky coverage is perhaps owing
to genuine field-to-field variations in counts.  To summarize, the work
presented here emphasises the importance of constucting source catalogs,
accounting for blending and noise bias, and making large area surveys in order to refine our
understanding of differential source counts at sub-mJy and mJy flux densities.

The main results presented in this paper concern the statistical properties of
the radio structure and polarization in sub-mJy sources compared to the mJy radio source
population. We have defined a complexity parameter $\chi$ as the ratio of the
total flux density of the source to the peak flux density in the high
resolution image, this parameter is a measure of the source complexity and the
departure of the source structure from an unresolved single compact
component. Additionally, we have defined a diffuseness parameter $\delta$ as
the ratio of total flux density to the sum of the flux density in
compact components, this parameter is a measure of the flux density in diffuse
emission components.  These measures of morphology have been computed for all
ATLBS sources. The points arising from an examination of
the sources above and below 1~mJy flux density are listed below.

\begin{enumerate}

\item 
In the low-resolution images made with $50\arcsec$ beam, the fraction of
extended sources rises from 15\% for the sub-mJy population to 28\% for 1--10~mJy
sources and to 70\% for 10--100~mJy sources. At this resolution, only 2\% of
sub-mJy sources are observed to have composite structure where as this
fraction rises to 15\% for 1--10~mJy sources and 50\% for 10--100~mJy sources.

\item
Less than 1\% of the sub-mJy ATLBS sources have been observed to have multiple
compact components.  However, about 10\% of 1--10~mJy sources have multiple
compact components and this fraction rises to 36\% in sources in the
10--100~mJy range.

\item
The median complexity $\chi$ for ATLBS sources is 1.28 and the median
diffuseness $\delta$
is 1.09.  20\% of ATLBS sources have $\chi$ exceeding 2.0, implying that a
fifth of the sources are doubles or triples or have more than half their flux
density in extended emission. 10\% of the ATLBS sources have $\delta$
exceeding 2.0, implying that a tenth of the sources have more than half their
flux density in diffuse emission.

\item
We observe no significant trend in median $\delta$ with flux density.
However, $\chi$ rises significantly with increasing flux density.  Whereas
only 8\% sub-mJy ATLBS sources have $\chi$ exceeding 2.0, 28\% of 1--10~mJy
sources have $\chi$ exceeding 2.0 and this fraction rises to 55\% for the
10--100~mJy sources.

\end{enumerate}

The sub-mJy
ATLBS sources, with 10\% sources having more than half the flux density in
extended emission, almost always have a single compact component, if present.
On the other hand, sources with higher flux density tend to have a greater
fraction of multiple compact components, although the fractional flux density
in the diffuse emission might be the same.  This is consistent with
population synthesis models for the differential source counts wherein the mJy
radio source population is dominated by the relatively powerful radio sources,
which are often of the hot-spot type with FR-{\sc ii} structure, and the sub-mJy sources
are dominated by the relatively lower power radio sources, which often
manifest the FR-{\sc i} structure with a single compact component.

We have computed the percentage integrated polarized intensity for the ATLBS
sources and examined their variation with total flux density.  As far as we
know, we have formulated herein a polarization bias correction for integrated
polarized emission for the first time.
We observe an increase in the percentage polarization in the ATLBS sources
with decreasing flux density.  The median percentage polarization is above
10\% in the sub-mJy sources and declines to a few percent in sources with
100~mJy flux density. We have been unable to find any correlation between the
percentage polarization and the source properties like size, complexity or the
fractional flux in the diffuse emission.  Since we
do observe a decrease in source complexity in fainter sources, it may be that
the increased percentage polarization in the fainter sources is owing to a
transition from FR-{\sc ii} dominated population at higher flux densities to
FR-{\sc i} dominated population. Radio source populations 
dominated by edge-darkened FR-{\sc i} jets or relaxed doubles and relict sources with
relatively homogeneous magnetic field orientation may suffer less
depolarization, due to averaging over the spatial extent of the source, as
compared to the FR-{\sc ii} sources that have more complex 
spatial structure in their field distributions. Alternately, the increased
fractional polarization in the fainter sources may be due to lower internal Faraday
depolarization if the fainter sources are at relatively higher
redshift and the emission frequency in the rest frame of the source is higher in
the case of the fainter sources. Higher resolution radio
imaging and redshift measurements of the optical
identifications of the ATLBS sources---both of which are currently
underway---might shed light on this issue.  

The Very Large Array (VLA) survey of the Chandra Deep Field South
\citep{ke08}, together with optical identifications of the radio sources
\citep{ma08}, suggest that the the sub-mJy radio sources above 0.08~mJy are
dominated by non-thermal emission associated with early type galaxies hosting
AGNs.  \citet{pa07} suggest that the dominant population is low-luminosity
AGNs of the FR-{\sc i} type.  The ATLBS study of radio
source morphology are consistent with this view.   

The ATLBS survey has indicated that the sub-mJy radio source population does
indeed have a non-negligible fraction of their integrated flux density in diffuse
emission. Therefore, it is vital that deep radio surveys be made with adequate 
surface brightness sensitivity to measure the total radio luminosity
associated with the AGNs, and accurately quantify any associated mechanical feedback.

The ATLBS survey regions are being re-observed using extended array
configurations of the ATCA to reconstruct source structures with higher angular
resolution, to confirm the findings presented in this paper and explore further
the radio structural properties of the sub-mJy radio 
source population.  Additionally, the wide fields are being mosaic observed in the
optical and IR bands using the CTIO Blanco Telescope and the
Anglo-Australian Telescope in a study of the host galaxies and their
environments. These observations and the consequent refinements in our
understanding of the evolution in extended radio sources will be the subject
of forthcoming papers. 

\section*{Acknowledgments}

The ATCA is part of the Australia Telescope, which is funded by the
Commonwealth of Australia for operation as a National Facility managed by CSIRO.

\end{document}